\documentclass[aps,prl,twocolumn,amsmath,amssymb,superscriptaddress,showkeys,showpacs]{revtex4}
\usepackage[dvips]{graphicx,color}
\usepackage[dvips]{hyperref}
\hypersetup{
  pdftitle={Determination of the total width of the eta-prime meson at COSY--11},
  pdfauthor={Eryk Miros{\l}aw Czerwi{\'n}ski, Pawe{\l} Moskal, Dieter Grzonka et al.},
  pdfkeywords={total width, eta-prime, missing mass},
  unicode=false,
  colorlinks=false,
}
\usepackage[english]{babel}
\newcommand{\ep}{$\eta^{\prime}$}

\newcommand{\epw}{$\Gamma_{\eta^{\prime}}$}

\newcommand{\ppep}{$pp\to pp\eta^{\prime}$}
\newcommand{\ppx}{$pp\to ppX$}
\newcommand{\cc}{\mbox{COSY--11}}
\begin{document}
\title{Determination of the total width of the $\eta^{\prime}$ meson}
\author{E.~Czerwi\'nski}\email[]{eryk.czerwinski@lnf.infn.it}
\affiliation{Institute of Physics, Jagiellonian University, PL-30-059 Cracow, Poland}
\affiliation{Institute for Nuclear Physics and J{\"u}lich Center for Hadron Physics,\\
                Research Center J{\"u}lich, D-52425 J{\"u}lich, Germany}
\author{P.~Moskal} 
\affiliation{Institute of Physics, Jagiellonian University, PL-30-059 Cracow, Poland}
\affiliation{Institute for Nuclear Physics and J{\"u}lich Center for Hadron Physics,\\
                Research Center J{\"u}lich, D-52425 J{\"u}lich, Germany}
\author{D.~Grzonka}
\affiliation{Institute for Nuclear Physics and J{\"u}lich Center for Hadron Physics,\\
                Research Center J{\"u}lich, D-52425 J{\"u}lich, Germany}
\author{R.~Czy\.zykiewicz}
\affiliation{Institute of Physics, Jagiellonian University, PL-30-059 Cracow, Poland}
\author{D.~Gil}
\affiliation{Institute of Physics, Jagiellonian University, PL-30-059 Cracow, Poland}
\author{B.~Kamys}
\affiliation{Institute of Physics, Jagiellonian University, PL-30-059 Cracow, Poland}
\author{A.~Khoukaz}
\affiliation{IKP, Westf\"alische Wilhelms-Universit\"at, D-48149 M\"unster, Germany}
\author{J.~Klaja}
\affiliation{Institute of Physics, Jagiellonian University, PL-30-059 Cracow, Poland}
\affiliation{Institute for Nuclear Physics and J{\"u}lich Center for Hadron Physics,\\
                Research Center J{\"u}lich, D-52425 J{\"u}lich, Germany}
\author{P.~Klaja}
\affiliation{Institute of Physics, Jagiellonian University, PL-30-059 Cracow, Poland}
\affiliation{Institute for Nuclear Physics and J{\"u}lich Center for Hadron Physics,\\
                Research Center J{\"u}lich, D-52425 J{\"u}lich, Germany}
\author{W.~Krzemie{\'n}}
\affiliation{Institute of Physics, Jagiellonian University, PL-30-059 Cracow, Poland}
\affiliation{Institute for Nuclear Physics and J{\"u}lich Center for Hadron Physics,\\
                Research Center J{\"u}lich, D-52425 J{\"u}lich, Germany}
\author{W.~Oelert}
\affiliation{Institute for Nuclear Physics and J{\"u}lich Center for Hadron Physics,\\
                Research Center J{\"u}lich, D-52425 J{\"u}lich, Germany}
\author{J.~Ritman}
\affiliation{Institute for Nuclear Physics and J{\"u}lich Center for Hadron Physics,\\
                Research Center J{\"u}lich, D-52425 J{\"u}lich, Germany}
\author{T.~Sefzick}
\affiliation{Institute for Nuclear Physics and J{\"u}lich Center for Hadron Physics,\\
                Research Center J{\"u}lich, D-52425 J{\"u}lich, Germany}
\author{M.~Siemaszko}
\affiliation{Institute of Physics, University of Silesia, PL-40-007 Katowice, Poland}
\author{M.~Silarski}
\affiliation{Institute of Physics, Jagiellonian University, PL-30-059 Cracow, Poland}
\author{J.~Smyrski}
\affiliation{Institute of Physics, Jagiellonian University, PL-30-059 Cracow, Poland}
\author{A.~T\"aschner}
\affiliation{IKP, Westf\"alische Wilhelms-Universit\"at, D-48149 M\"unster, Germany}
\author{M.~Wolke}
\affiliation{Institute for Nuclear Physics and J{\"u}lich Center for Hadron Physics,\\
                Research Center J{\"u}lich, D-52425 J{\"u}lich, Germany}
\affiliation{Department of Physics and Astronomy, Uppsala University, SE-751 20 Uppsala, Sweden}
\author{P.~W\"ustner}
\affiliation{Institute for Nuclear Physics and J{\"u}lich Center for Hadron Physics,\\
                Research Center J{\"u}lich, D-52425 J{\"u}lich, Germany}
\author{J.~Zdebik}
\affiliation{Institute of Physics, Jagiellonian University, PL-30-059 Cracow, Poland}
\author{M.~Zieli\'nski}
\affiliation{Institute of Physics, Jagiellonian University, PL-30-059 Cracow, Poland}
\author{W.~Zipper}
\affiliation{Institute of Physics, University of Silesia, PL-40-007 Katowice, Poland}
\date{\today}
\begin{abstract}
Taking advantage of both the low-emittance proton beam of
the cooler synchrotron COSY and the  high momentum precision of the COSY-11 detector system,
the mass distribution of the $\eta^{\prime}$ meson was measured with a resolution of 0.33 MeV/c$^2$ (FWHM),
improving the experimental mass resolution by almost an order of magnitude  with respect to
previous results.
Based on the sample of more than 2300 reconstructed $pp\to pp\eta^{\prime}$ events
the total width of the \ep\ meson  was determined to be
$\Gamma_{\eta^{\prime}}~=~0.226~\pm~0.017(\textrm{stat.})~\pm~0.014(\textrm{syst.})$~MeV/c$^2$.
\end{abstract}
\pacs{13.60.Le, 14.40.Be, 14.70.Dj}
\keywords{total width, eta-prime meson, missing mass}
\maketitle
In this Letter, we report on the measurement of the mass distribution 
of the \ep\ meson carried out with a
resolution of a fraction of MeV/c$^2$. This accuracy
was obtained by using the low-emittance proton-beam of 
the cooler synchrotron COSY~\cite{Maier} 
and the  high momentum resolution of the COSY-11 detector 
system~\cite{Brauksiepe,Moskal22},  and it
is nearly an order of magnitude more precise than
previous results.

In the latest review by the Particle Data Group (PDG)~\cite{pdg}, two values 
for the total width of the \ep\ meson are given.  One of these values, (0.30$\pm$0.09)~MeV/c$^2$, results from the average of 
two measurements~\cite{Wurzinger,Binnie}, though
only in one of these experiments was \epw\ extracted directly
based on the mass distribution~\cite{Binnie}. 
The second value (0.205$\pm$0.015)~MeV/c$^2$, recommended by the PDG, 
is determined by fit 
to altogether 51 measurements of 
partial widths, branching ratios, and
combinations of particle widths obtained from integrated cross sections~\cite{pdg}.
The result of the fit is strongly correlated with the value of 
the partial width $\Gamma(\eta^{\prime}\to\gamma\gamma)$,
which causes serious difficulties when the total and the partial width have to be used at the same time, 
like e.g. in studies of the gluonium content of the \ep\
meson~\cite{Biagio,Ambrosino2,Biagio2}.

The partial width 
of the $\eta^{\prime}\to\gamma\gamma$ channel can be extracted from the $e^+e^- \to e^+e^-\eta^{\prime}$
cross sections without knowledge of the \epw~\cite{Acciarri,Karch,Williams}, yet its derivation is model-dependent
due to the need to incorporate a form factor
which describes the spatial distribution of the electric charge in the $\eta^{\prime}$ meson.
For the derivation of partial widths of all other decay channels, the knowledge of \epw\ is mandatory.
At present it is the inaccuracy  of \epw\  which limits 
investigations of many interesting physics issues, such
as, for example, the quark mass difference $m_{d}-m_{u}$~\cite{Borasoy2,Zielinski,Kupsc}, 
isospin breaking in QCD~\cite{Borasoy,Borasoy2}, or 
the box anomaly of QCD~\cite{Nissler}.
This is because the branching ratios of the \ep\ meson decay channels are typically
known with a relative precision of more than 
an order of magnitude better than the present
accuracy with which \epw\ is extracted~\cite{pdg}.

The signal of the \ep\ meson production observed in previous
experiments~\cite{Moskal4,Moskal8,Khoukaz,Hibou,Wurzinger,Balestra,Binnie,PKlaja} with mass resolutions
poorer than $\sim$1~MeV/c$^2$ do not \mbox{\emph{a priori}} exclude the possibility 
that some structure in the mass distribution of the \ep\ meson would be visible at higher precision.
Extractions of the \ep\ width (\epw) were performed
under the assumption that the \ep\ meson is
a single state. This, however, must not necessarily be
the case~\cite{AguilarBenitez,Lasinski} 
if there is a significant glue contribution in the wave function of this meson~\cite{Bass}.
The precision achieved with the COSY-11 facility enabled us for the first time to determine 
the mass spectrum of the \ep\ meson with a resolution comparable
to its total width of $\sim$0.2~MeV/c$^2$~\cite{pdg}.

The experiment, reported in this Letter, was performed
in the Research Centre J{\"u}lich.
The value of \epw\ was established directly from the measurement of the mass distribution of the \ep\ meson,
produced via the \ppep\ reaction.
The momentum of the COSY beam and the dedicated zero degree COSY-11 facility
enabled the measurement at an
excess energy  of only a fraction of an MeV above the kinematic threshold  
for the \ep\ meson production. 
This was the most decisive factor in minimizing uncertainties of the 
missing-mass determination,
since at threshold the partial derivative of the missing mass
with respect to the outgoing proton momentum tends to zero.
In addition, close to threshold
the signal-to-background ratio increases due to the more rapid reduction of the phase space  for
multimeson production than for~the~\ep.

\begin{figure}[!b]
\vspace{-10mm}
\begin{flushright}
    \includegraphics[width=0.39\textwidth]{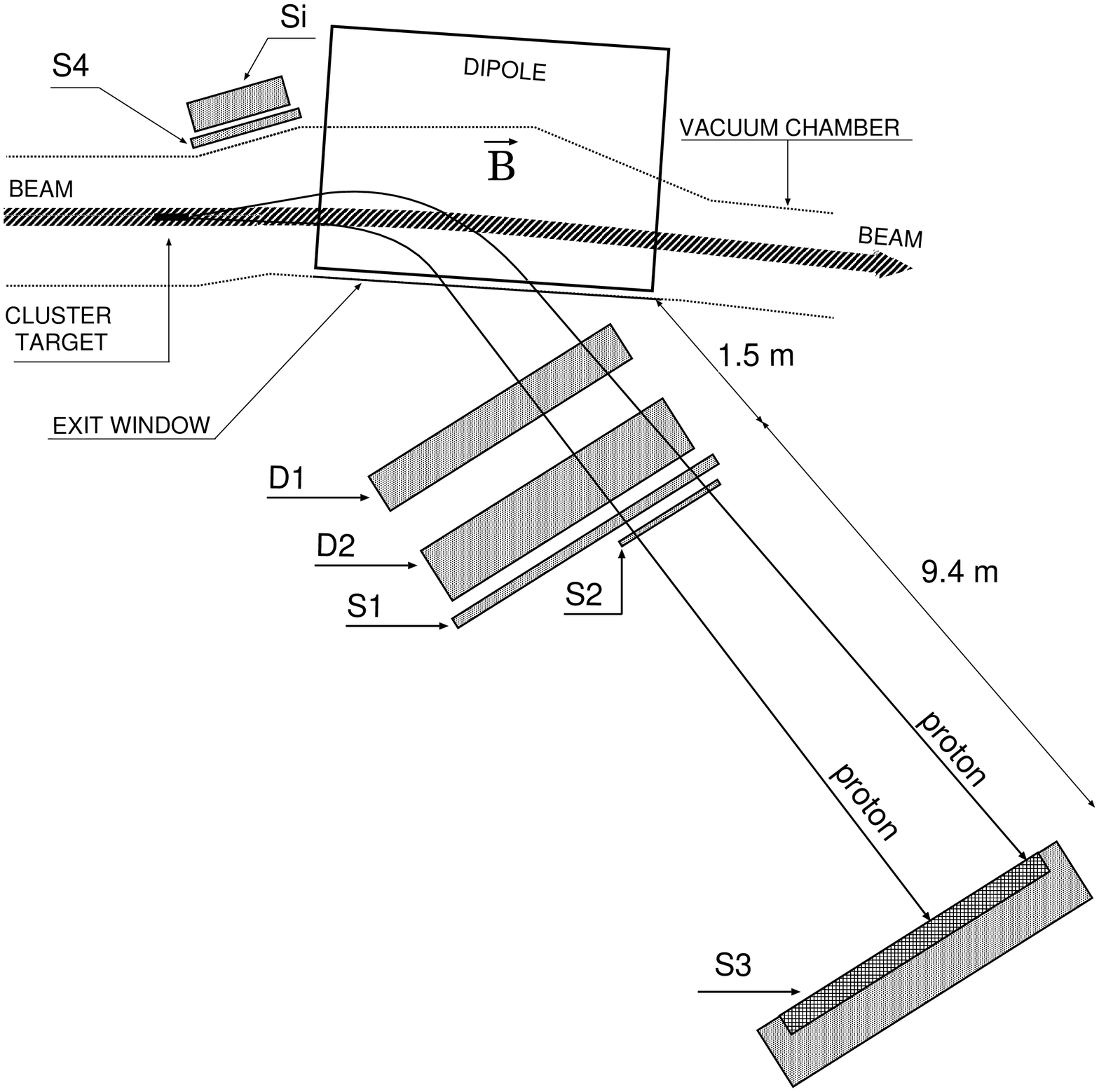}
\end{flushright}
\vspace{-40mm}
\begin{flushleft}
    \includegraphics[width=0.19\textwidth]{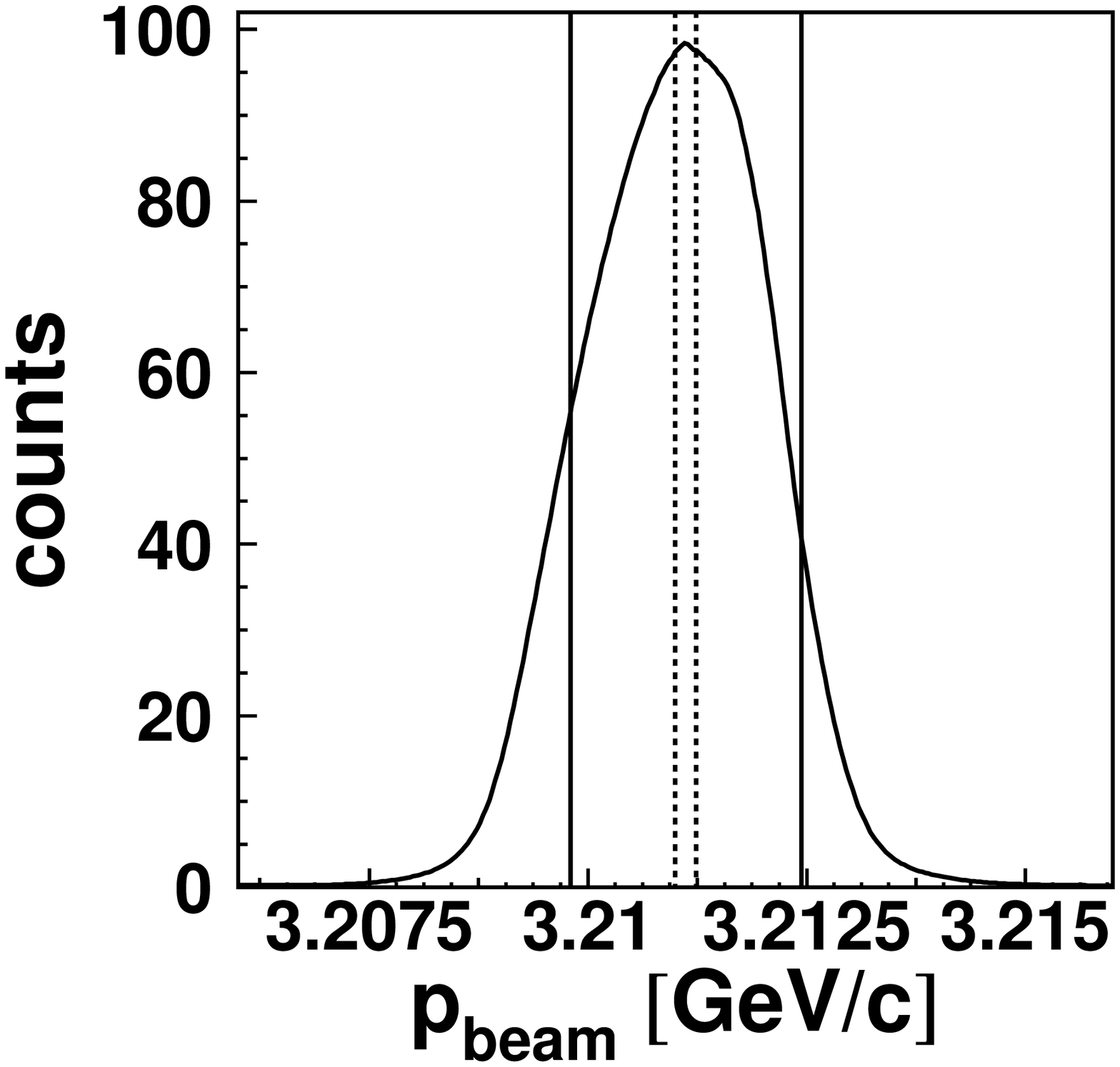}
\end{flushleft}
\vspace{-5.7mm}
 \caption{
         Schematic view of the \cc\ detector setup (top view).
         S1, S2, S3 and S4 denote scintillator detectors, D1 and D2 indicate drift chambers and  Si stands 
         for the silicon-pad detector.
         Left corner: Momentum spectrum for the measurement 
         with the nominal beam momentum of 3211~MeV/c. As an example, 
         the effective spread of the beam momentum due to the dispersion
         is shown as for a target width of 1~mm (dashed line) and 1~cm (solid line).
         }
 \label{c11}
\vspace{-7mm}
\end{figure}

In order to control systematic uncertainties,
the measurement was carried out at five different
beam momenta, which were nominally 3211, 3213, 3214, 3218, and 3224~MeV/c.
The cooled beam of protons~\cite{VanDerMeer,Prasuhn} circulated in the ring of the cooler synchrotron COSY
through a stream of the hydrogen cluster target~\cite{Dombrowski}.
In the magnetic field of the COSY dipole,
the final state protons from the \ppep\ reaction 
were bent more than the beam protons and were measured by means of the \cc\ detector
shown schematically in Fig.~\ref{c11}.
The momentum vectors of the outgoing protons were reconstructed based on the 
bending of their trajectories in the magnetic field between the center of the 
reaction region and the tracks measured in the drift chambers (D1 and D2). 
In addition, the velocities of the two protons were 
determined from their time of flight measured between 
the scintillator detectors S1 and S3.
The independent determination of momentum and velocity
enables particle identification to be made via its invariant mass. Since the momentum is
reconstructed more precisely than the velocity, after the identification,
the energy of the particle
is derived from its known mass and momentum.
The standard technique for monitoring the beam momentum at the COSY accelerator is through the measurement of the
frequency distribution of the circulating beam. Such a distribution can be transformed to the momentum coordinate
by using the values of the accelerator settings~\cite{Prasuhn}.
As an example, a spectrum for the lowest beam energy used in the experiment
is presented in the left corner of Fig.~\ref{c11}.
The beam momentum distribution is smooth
and its spread is equal to 2.5~MeV/c (FWHM).
However, due to the position of the \cc\ target system in a bending section of the COSY ring
in a dispersive region, the effective spread of the beam (the momentum range \emph{seen} by the target)
is smaller~\cite{ErykPhD}.
The \ep\ meson was not registered but instead it was identified 
by using the missing-mass technique.
The precision of the determination of the size and position of the target stream
influences the accuracy of the reconstruction of the momentum of the outgoing particles 
and the accuracy of the determination of the momentum spread 
of beam protons interacting with the target. 
As a compromise between accuracy and statistics, the transverse size of the target stream 
was reduced to 0.9~mm which is significantly less than the horizontal spread of the COSY beam.
Therefore, the momentum spread of the interacting protons is defined by
the momentum dispersion at the target region and by the size of the target stream. 
The size and position of the target stream, being crucial for the analysis,
were monitored by two independent methods.
The first was based on the measurement of the momentum distribution of elastically
scattered protons~\cite{Moskal5},
while the second was a direct measurement of the target geometry by
mechanically scanning the target stream position above and below the target area
from time to time.
A diagnostic unit with several wires was rotated through the target stream and the
pressure in the cluster beam dump was measured as a function of the wire position.
When parts of the target stream are blocked by a wire, the pressure decreases proportional
to the blocked area.
Therefore variations of
the pressure allowed the monitoring of
the size and alignment of the target stream during the experiment.
The results of the two methods are in good agreement and the achieved precision is $\pm$0.05 and $\pm$0.01~mm for the size and alignment, respectively. 
The momentum distributions of the elastically scattered protons were used not only for monitoring
the relative geometrical settings of the target, but also for the exact positions of the dipole field and the drift chambers~\cite{ErykPhD}.

\begin{figure}[!b]
 \vspace{-5mm}
  \begin{center}
    \includegraphics[width=0.47\textwidth]{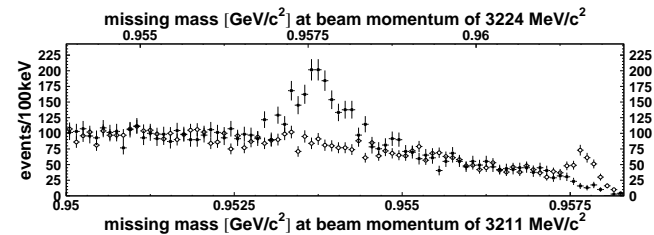}
  \end{center}
 \vspace{-7mm}
 \caption{
          Missing-mass spectra for the \ppx\ reaction
          determined at beam momenta of 3211 (open points) and 3224~MeV/c (filled points).
          The filled points were shifted to the kinematic limit and normalized to the open points.
         }
 \label{smoothbcg}
 \vspace{-5mm}
\end{figure}

As the next step of the analysis, the missing-mass spectra 
were determined in order (i) to distinguish between the signal and background,
(ii) to evaluate the absolute beam momenta, and finally (iii) to
extract the width of the \ep\ meson.
The spectra for the highest and lowest excess energies are shown in Fig~\ref{smoothbcg}.
It is important to stress that the background
distribution is smooth in the whole range studied,
and that the signal from \ep\ meson production shows up clearly.
The figure illustrates that the spectrum 
at one energy can be used as a good estimate of the background to the spectra at the other energies. 
The method for the background subtraction is based on the observation that 
the shape of the multipion
mass distribution does not change when the excess energy for
the \ppep\ reaction varies by a few MeV,
which is small compared to the total available energy of about 500~MeV~\cite{Moskal4}.
The systematic error in the changes of the shape due to the method applied
was estimated to be less than 1\% even for shifts many times  
larger than the energy range relevant in these measurements~\cite{Moskal7}.
In order to decrease the influence of the statistical fluctuation, the background
for a given energy was taken from a second-order polynomial fit to the data
at a different energy, which was shifted and normalized to the data of interest~\cite{ErykPhD}.

Owing to the large statistics of the momentum distributions for elastically scattered protons,
relative differences between excess energies were determined 
with a negligible statistical error from the sizes of the kinematic ellipses~\cite{ErykPhD}.
Next, the absolute values of the excess energies were derived by comparing the
position of the mean of the missing-mass peak for data closest to the threshold
with the empirical value of the mass of the \ep\ meson~\cite{pdg}.
The "true" values of the beam momenta thus determined are
3210.7, 3212.6, 3213.5, 3217.2, and 3223.4~MeV/c, 
corresponding to excess energies 
of 0.8, 1.4, 1.7, 2.8, and 4.8~MeV, respectively. 
The accuracy of the
beam momentum determination amounts to $\pm0.2$~MeV/c and is predominantly due to the
uncertainty of the \ep\ mass (957.78~$\pm$~0.06)~MeV/c$^2$~\cite{pdg}. 
The systematically lower values of the true beam momenta 
of about 0.5~MeV/c are consistent with previous
experience at COSY where the real beam momentum 
was always smaller than the nominal value~\cite{Moskal6,Smyrski3}.

\begin{figure}[!b]
  \vspace{-1mm}
  \begin{minipage}[c]{0pt}
    \raisebox{0pt}[0pt][0pt]{
      \mbox{
        \raisebox{-2.1cm}{\hspace{-3.30cm}\footnotesize{\textsf{Q=4.8~MeV}}}
        \raisebox{-2.1cm}{\hspace{ 2.60cm}\footnotesize{\textsf{Q=2.8~MeV}}}
      }
    }
  \end{minipage}
  \begin{minipage}[c]{0pt}
    \raisebox{0pt}[0pt][0pt]{
      \mbox{
        \raisebox{-5.60cm}{\hspace{-3.40cm}\footnotesize{\textsf{Q=1.7~MeV}}}
        \raisebox{-5.60cm}{\hspace{ 2.55cm}\footnotesize{\textsf{Q=1.4~MeV}}}
      }
    }
  \end{minipage}
  \begin{minipage}[c]{0pt}
    \raisebox{0pt}[0pt][0pt]{
      \mbox{
        \raisebox{-9.10cm}{\hspace{-5.00cm}\footnotesize{\textsf{Q=0.8~MeV}}}
      }
    }
  \end{minipage}
  \begin{center}
  \vspace{-15mm}
    \includegraphics[width=0.23\textwidth]{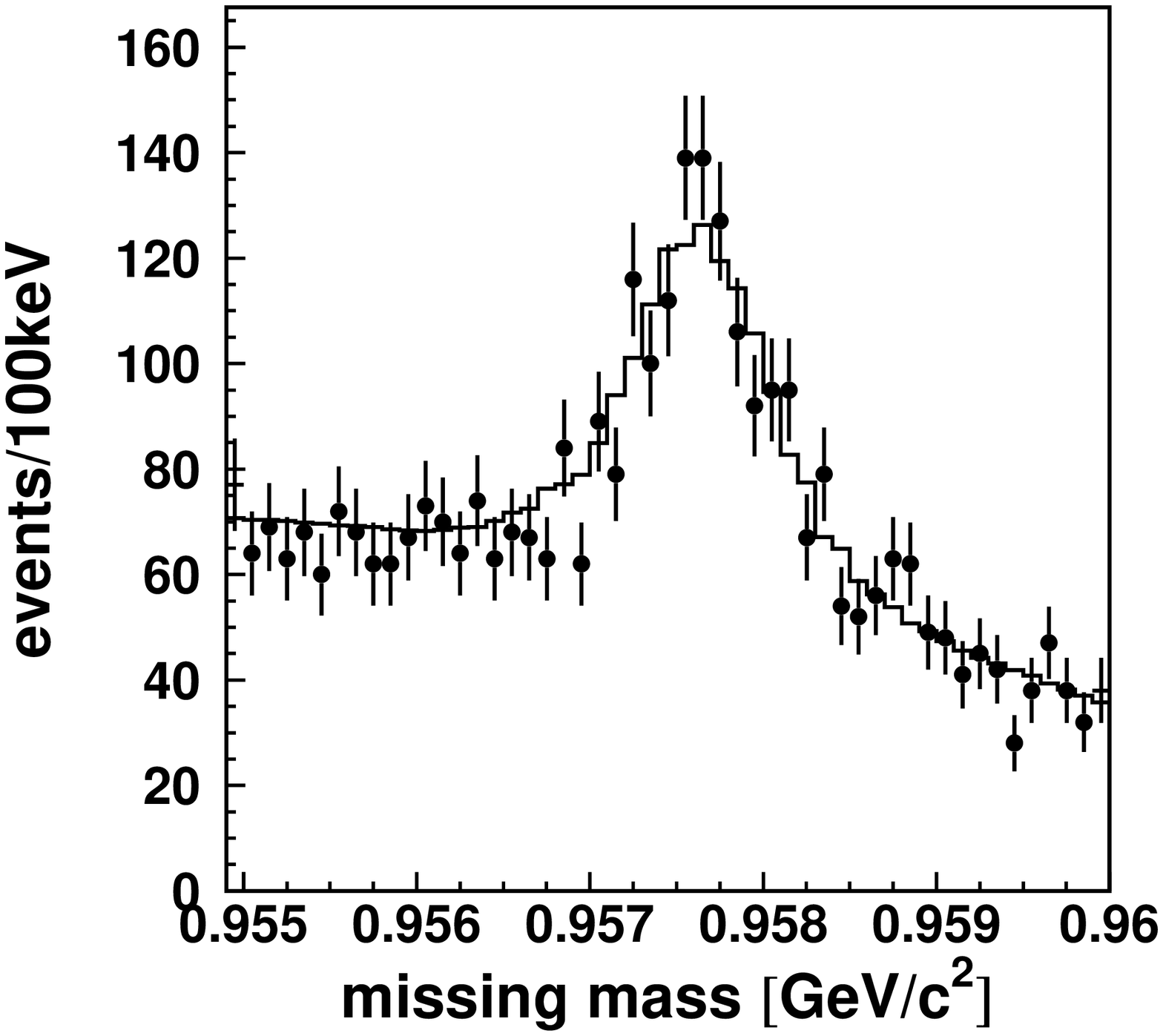}
  \vspace{-7mm}
    \includegraphics[width=0.23\textwidth]{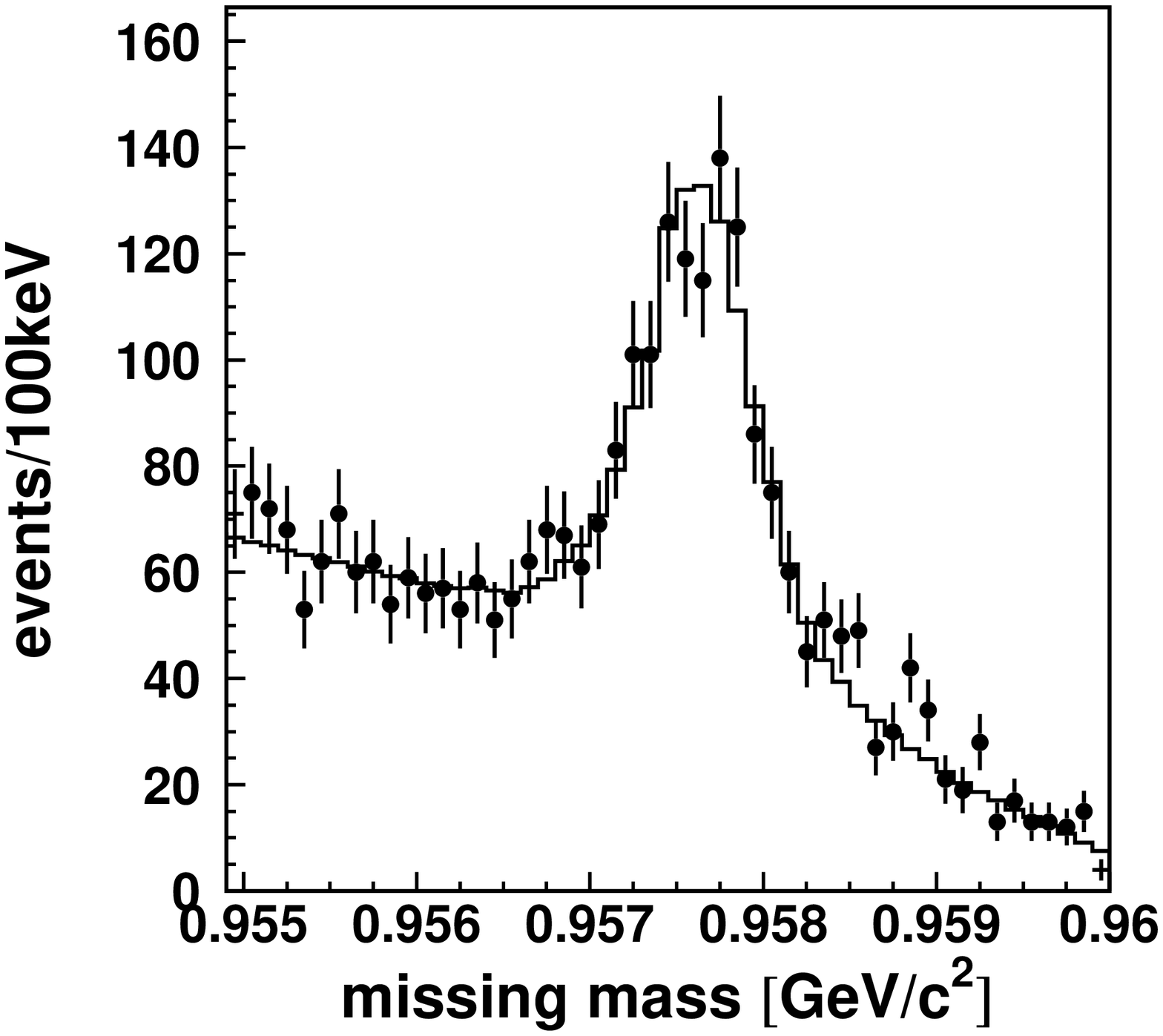}
    \includegraphics[width=0.23\textwidth]{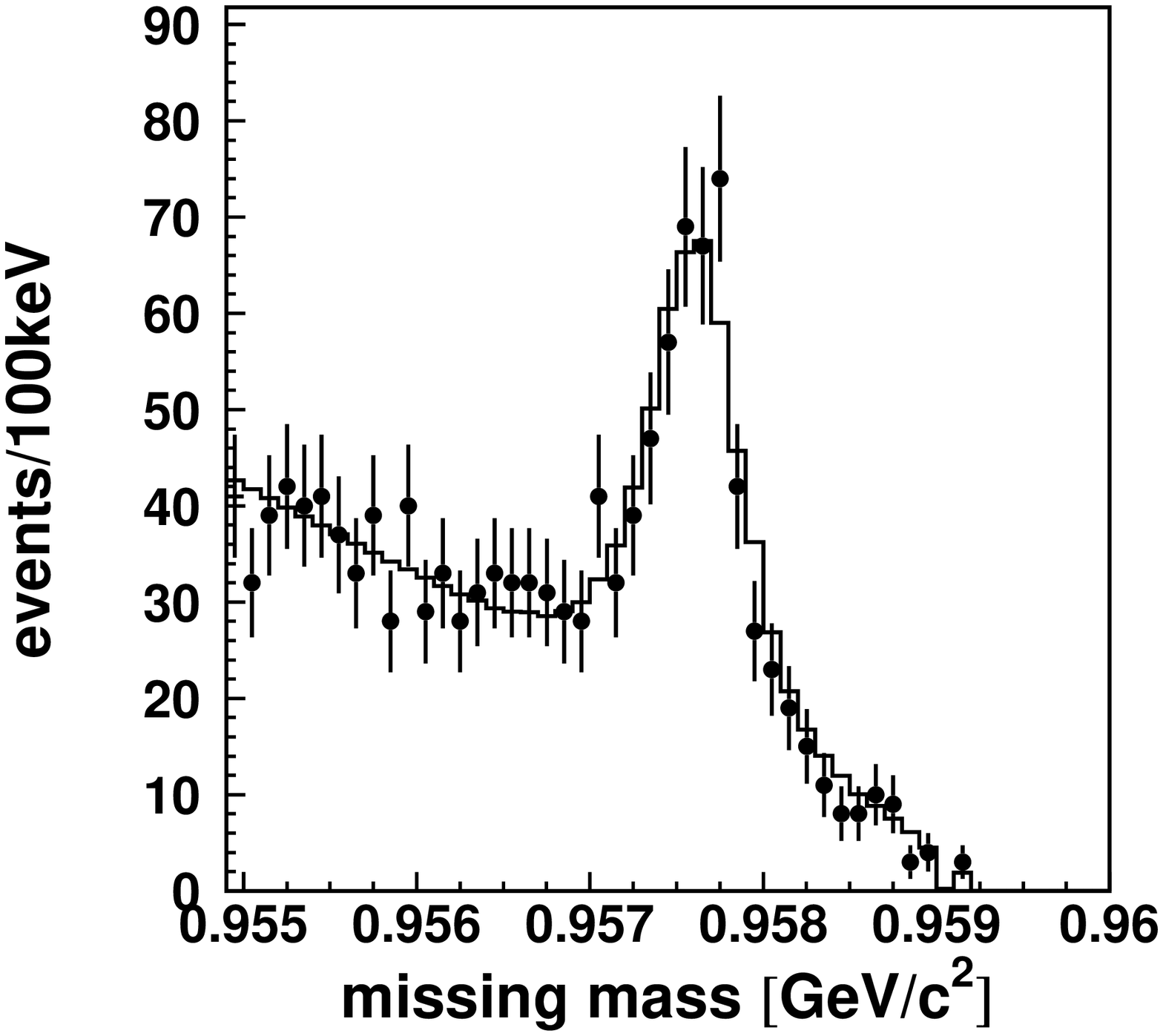}
  \vspace{-7mm}
    \includegraphics[width=0.23\textwidth]{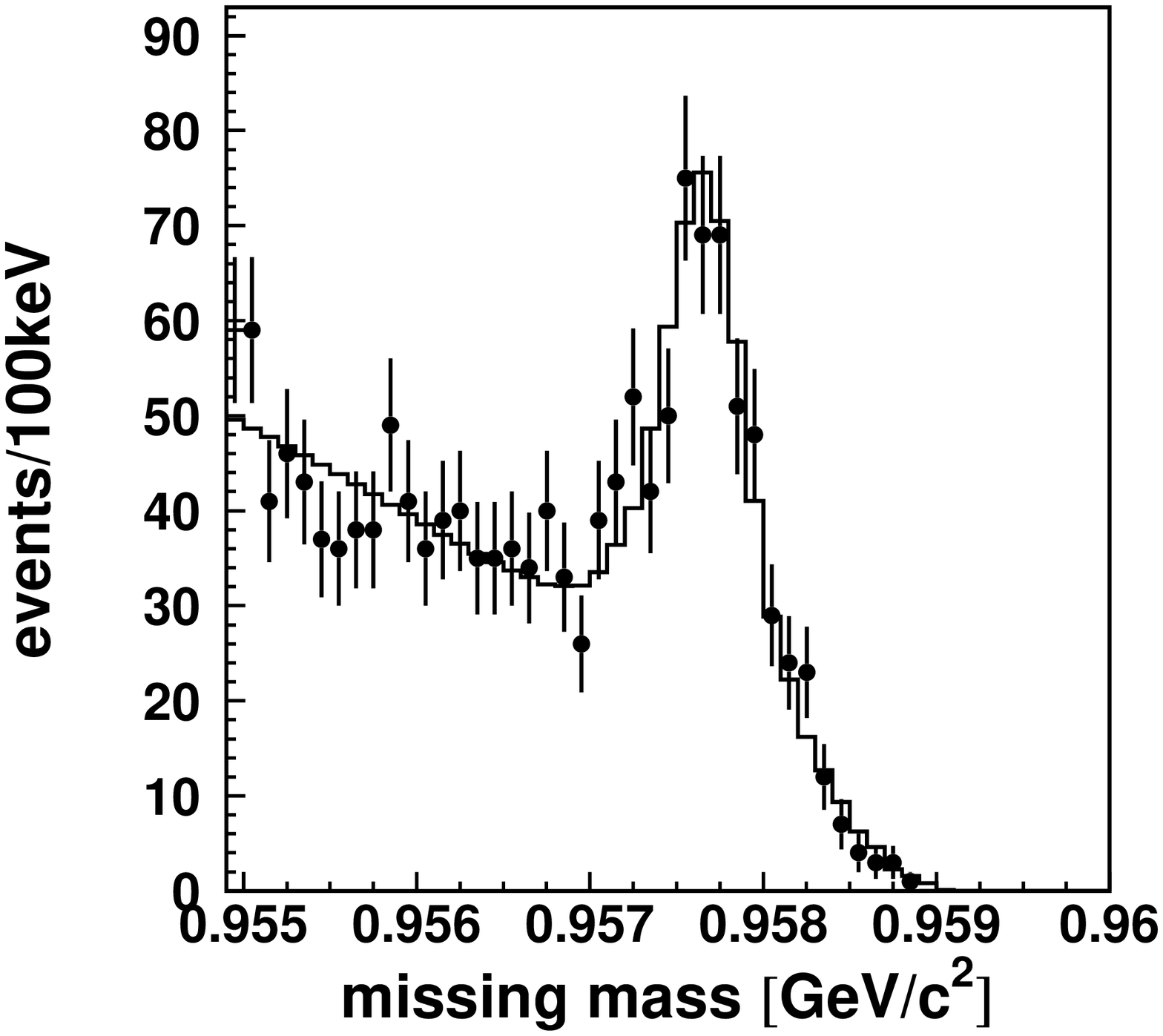}
    \includegraphics[width=0.23\textwidth]{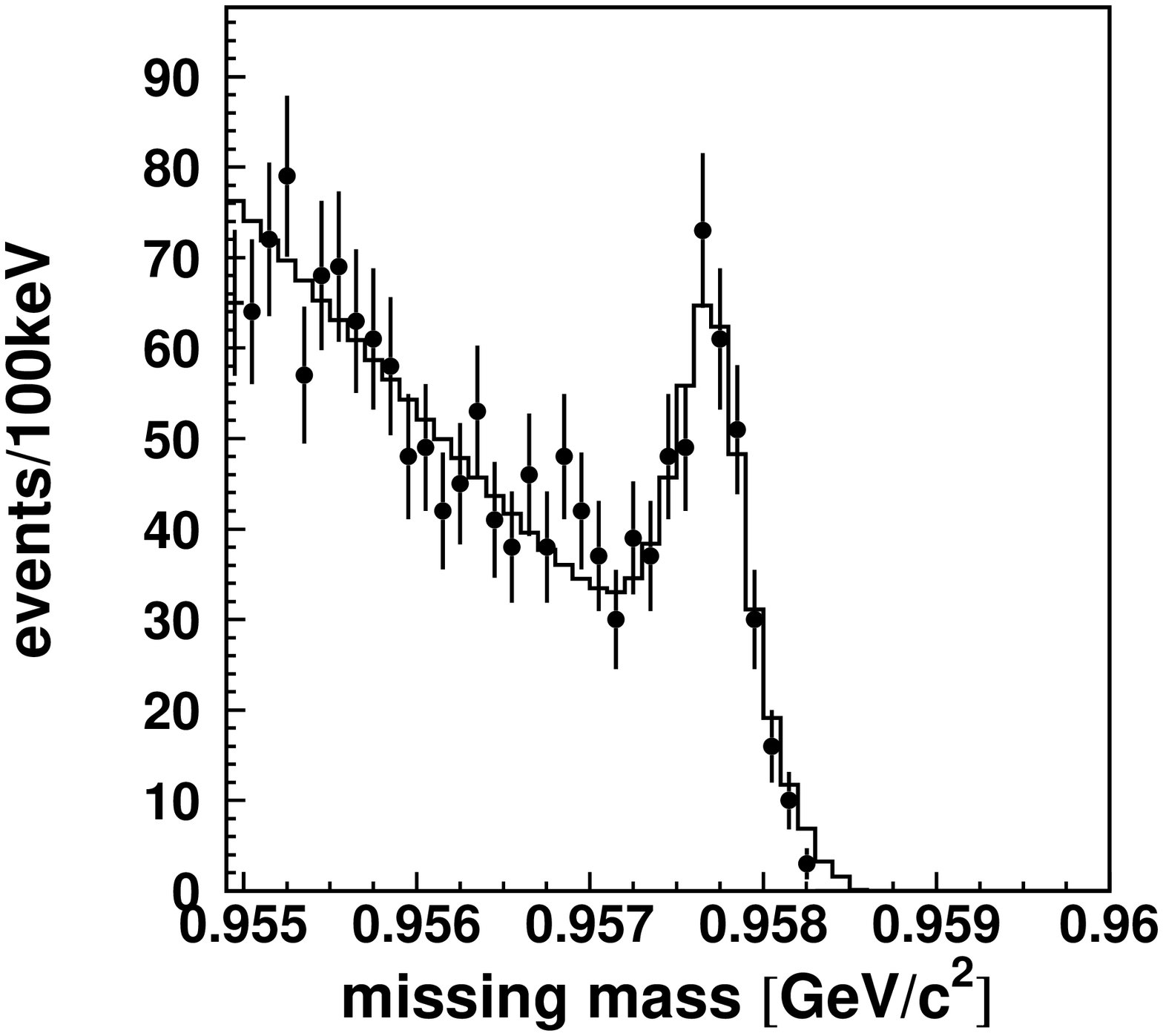}
  \vspace{-5mm}
    \includegraphics[width=0.23\textwidth]{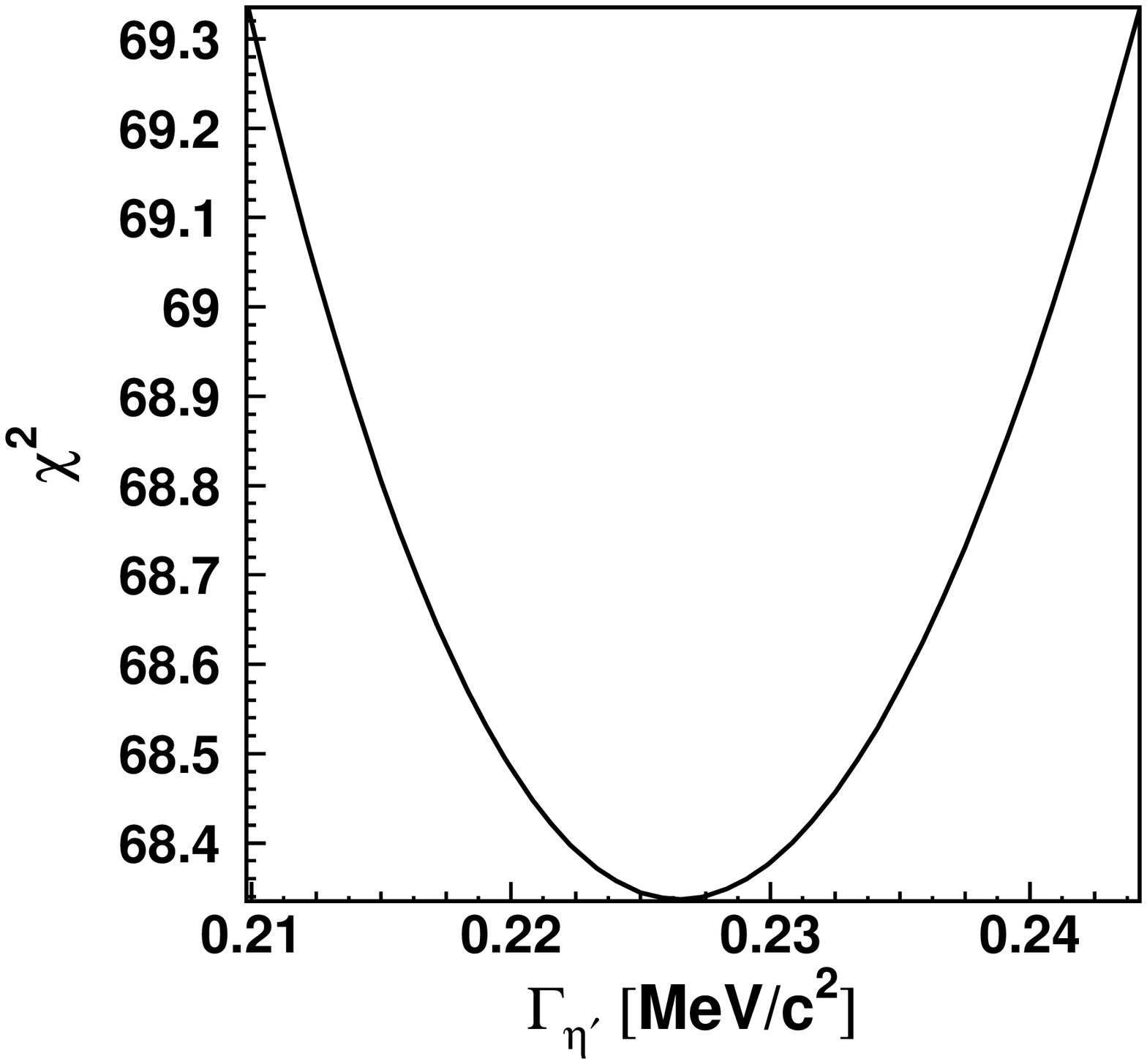}
  \end{center}
  \vspace{-2mm}
 \caption{
         The missing-mass spectra for the \ppx\ reaction.
         The \ep\ meson signal is clearly visible. 
         The experimental data are presented
         as points, while in each plot the line corresponds to the sum of the Monte Carlo generated signal
         for the \ppep\ reaction with $\Gamma_{\eta^{\prime}}=0.226$~MeV/c$^2$ and the
         background obtained from another energy.
         The plot at the bottom right of the figure presents  $\chi^2$ as a function of the~\epw.
         The minimum value of the $\chi^2$ divided by
         the number of degrees of freedom amounts to 0.96.
         }
 \label{mmbcgfit}
\end{figure}

In order to derive the value of the \ep\ width, 
the experimental missing-mass spectra were compared with distributions
simulated with different values of \epw.
In the simulations based on the GEANT3 packages~\cite{geant3www}, the response of the COSY-11 detector system 
to the  \ppep\ reaction was generated,
taking into account the geometry and material composition, 
as well as relevant resolutions of the \cc\ detector components,
including the size of the target stream, the spatial and momentum spread of the beam,
and also all known physical processes such as multiple scattering and nuclear reactions. 
In the simulations of the mass distribution, a Breit-Wigner formula for the \ep\ meson was used.
Afterwards the generated events were analyzed in the same way as the experimental data,
and sets of missing-mass spectra were reconstructed for the values of \epw\ ranging from 0.14 to 0.38~MeV/c$^2$.
Finally, the sum of the experimental background and 
the Monte Carlo missing-mass spectra 
for the \ppep\ reaction 
was fitted to the experimental data.
The normalization factor of the $\eta^{\prime}$ signal was the only free parameter in this fit.
The result of the fit is shown by solid lines in Fig.~\ref{mmbcgfit}.
The decrease of the width of the
missing-mass with decreasing excess energy is a kinematical effect
reflecting the propagation of errors of momenta involved in the missing mass calculations~\cite{ErykMgr}.
The simulations reproduce very well the change
of the signal width with excess energy
and thus validate the correctness of the established detector and target characteristics.
The lower-right panel of Fig.~\ref{mmbcgfit} presents the dependence of the $\chi^2$ on the  \epw\ value. 
The minimum of $\chi^2$ is at $\Gamma_{\eta^{\prime}}~=~0.226$~MeV/c$^2$ and the
$1\sigma$ statistical error 
is equal to $\pm0.017$~MeV/c$^2$.

\begin{figure}[!t]
  \begin{center}
    \includegraphics[width=0.45\textwidth]{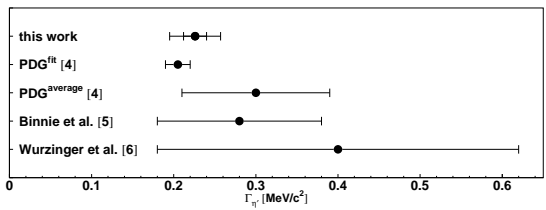}
  \end{center}
 \vspace{-7mm}
 \caption{
          Comparison of available values of \epw.
         }
 \label{gammas}
 \vspace{-5mm}
\end{figure}

The systematic error was estimated by studying the sensitivity of the result to the 
variation of parameters describing the experimental conditions 
in the analysis and in the simulation~\cite{ErykPhD}.
The contributions to the systematic error 
are (i) the target position~($\pm 0.006$~MeV/c$^2$) 
and size ($\pm 0.002$~MeV/c$^2$),
(ii) the position and orientation of the drift chambers~($\pm 0.001$~MeV/c$^2$),
(iii) the map of the magnetic field~($\pm 0.007$~MeV/c$^2$), 
and (iv) the absolute beam momentum determination~($\pm 0.003$~MeV/c$^2$). These values
were estimated as the difference between the derived result
of the \epw\ and the \epw\ values
established by changing in the analysis and simulations a particular parameter by its error.
The systematic error due to the method of the background subtraction~($\pm 0.006$~MeV/c$^2$) 
was established as the maximum difference
between \epw\ values determined when using experimental background shapes from different
energies. The uncertainty due to the bin width~($\pm 0.004$~MeV/c$^2$) was estimated 
by changing the width of bins
in the range from 0.1 to 0.04~MeV/c$^2$.
Furthermore, the sensitivity of the result to the range
of the missing-mass values used for the fit
($\pm 0.005$~MeV/c$^2$) was estimated  by enlarging the mass range by  seven bins
on each side of the peak.  The inaccuracy due to the model applied in the simulations for the 
proton-proton final state interaction ($\pm 0.003$~MeV/c$^2$)
was estimated conservatively as a differences in results
determined when using parameterization
of the proton-proton S-wave interaction~\cite{Naisse,Moskal9} and when neglecting the final-state interactions. 
Finally, the total systematic error was estimated
as the quadratic sum of the nine independent contributions mentioned above
and is $0.014$~MeV/c$^2$. Our final result is compared with earlier width determinations in Fig.~\ref{gammas}.

In summary, the mass distribution of the \ep\ meson has been measured 
with an experimental resolution of FWHM~=~0.33~MeV/c$^2$.
The \ep\ meson was created in the \ppep\ reaction
close to the kinematic threshold
by using the low-emittance proton beam of the cooler synchrotron COSY
incident on a stream of hydrogen clusters. The outgoing protons were
detected by using the COSY-11 facility.
The total width of the \ep\ meson was extracted from the missing-mass spectra
and amounts to $\Gamma_{\eta^{\prime}}=0.226\pm0.017(\textrm{stat.})\pm0.014(\textrm{syst.})$~MeV/c$^2$.
The result does not depend on knowing any of the branching ratios or partial decay widths. 
The extracted \epw\ value is in agreement 
with both previous direct determinations of this value
($\Gamma_{\eta^{\prime}}=0.28\pm0.10$~MeV/c$^2$~\cite{Binnie} and $\Gamma_{\eta^{\prime}}=0.40\pm0.22$~MeV/c$^2$~\cite{Wurzinger}). 
The achieved accuracy is similar to that obtained by the PDG from a fit to 51
measurements of branching ratios and cross sections ($\Gamma_{\eta^{\prime}}=0.204\pm0.015$~MeV/c$^2$)~\cite{pdg}.

Finally, it is worth noting that the achieved mass resolution 
is of the same order as the total width of the \ep\ meson itself, 
thereby excluding the possibility of a substructure in the \ep\ signal at this level.
\begin{acknowledgments}
We are thankful to Professor Colin Wilkin for his valuable comments and suggestions 
of improvements of this manuscript.
The work was partially supported by the European Commission 
through the \emph{Research Infrastructures} action of the \emph{Capacities} Program.
Call: FP7-INFRASTRUCTURES-2008-1, Grant Agreement No. 227431, by the PrimeNet,
by the Polish Ministry of Science and Higher Education through Grants No. 1253/B/H03/2009/36 and No. 1202/DFG/2007/03,
by the German Research Foundation (DFG), by the FFE
grants from the Research Center J{\"u}lich, and by the virtual institute \emph{Spin and strong
QCD} (VH-VP-231).
\end{acknowledgments}
\vspace{-4.6mm}

\end{document}